\documentclass{iopart}             
\usepackage{graphicx}
\usepackage{epsfig,epsf,graphics,psfig,lscape,url}
\usepackage[square, comma, sort&compress]{natbib}
\begin{document}
\setlength{\arraycolsep}{2.5pt}             
\jl{2}
%
%
%
\def\etal{{\it et al~}}
\def\newblock{\hskip .11em plus .33em minus .07em}
%
%
%
%
%
%
\title[Photoionization of Se$^+$]{Experimental Photoionization
Cross-Section Measurements in the Ground and Metastable State Threshold Region of Se$^+$}

\author{N C Sterling$^{1, 2}$,
  D A Esteves$^{3, 4}$,
  R C Bilodeau$^{4, 5}$,
  A L D Kilcoyne$^{4}$,
  E C Red$^{4}$,
  R A Phaneuf$^{3}$,
  and A Aguilar$^{4}$
  }

\address{$^{1}$Department of Physics and Astronomy, Michigan State University, 3248
               Biomedical Physical Sciences, East Lansing, MI 48824-2320, USA}

\address{$^{2}$NSF Astronomy and Astrophysics Postdoctoral Fellow}

\address{$^{3}$Department of Physics, MS 220, University of Nevada, Reno, NV 89557, USA}

\address{$^{4}$Advanced Light Source, Lawrence Berkeley National Laboratory, Berkeley,
               CA 94720, USA}

\address{$^{5}$Department of Physics, Western Michigan University, Kalamazoo, MI 49008, USA}

\begin{abstract}
Absolute photoionization cross-section measurements are reported for Se$^+$ in
the photon energy range 18.0--31.0~eV, which spans the ionization
thresholds of the $^4S_{3/2}$ ground state and the low-lying
$^2P_{3/2,1/2}$  and $^2D_{5/2,3/2}$ metastable states. The
measurements were performed using the Advanced Light Source
synchrotron radiation facility. Strong photoexcitation-autoionization resonances
due to 4$p$$\rightarrow$$nd$ transitions are seen in the cross-section spectrum
and identified with a quantum-defect analysis.
\end{abstract}
%
%
%

\vspace{0.25cm} {\begin{flushleft} Short title: Photoionization
of Se$^{+}$
\\

\vspace{0.25cm}
J. Phys. B: At. Mol. Opt. Phys: \today
\end{flushleft}}
\maketitle
%
%
%
%

%
\section{Introduction}
Photoionization is an important process in determining the
ionization balance and hence the abundances of elements in
astrophysical nebulae.  In the last few years, it has become possible to
detect neutron(\emph{n})-capture elements (atomic
number $Z>30$) in a large number of ionized nebulae
\cite{sterling07,sterling08}.  These elements are produced by
slow or rapid \emph{n}-capture nucleosynthesis (the
``\emph{s}-process'' and ``\emph{r}-process,'' respectively).
Measuring the abundances of these elements can reveal
their dominant production sites in the Universe, as well as
details of stellar structure, mixing and nucleosynthesis
\cite{smith90,wally97,busso99,trav04,herwig05,sneden08,karakas09}. These
astrophysical observations provide an impetus to determine the
photoionization and recombination properties of
\emph{n}-capture elements.

Various \emph{n}-capture elements have been detected in the spectra of planetary
nebulae
\cite{pequignot94,sharpee07,sterling08,sterling09},
the photoionized ejecta of
evolved low- and intermediate-mass stars (1--8 solar masses).  Planetary nebula progenitor stars may
experience \emph{s}-process nucleosynthesis \cite{busso99, stran06, karakas09,karakas10},
in which case their nebulae
will exhibit enhanced abundances of trans-iron elements.  The level of \emph{s}-process
enrichment for individual elements is strongly sensitive to the physical conditions
and mixing processes in
the stellar interior \cite{busso99, herwig05, karakas09}.

The principal difficulty in studying \emph{s}-process
enrichments in planetary nebulae is the large uncertainties (factors of 2 to 3)
of \emph{n}-capture element abundances derived from the
observational data.  There are two root causes for these
uncertainties.  First, due in part to their low cosmic abundances,
only one or two ions of a given \emph{n}-capture element can be
detected in individual planetary nebulae.  To derive elemental abundances,
corrections must be applied for the abundances of unobserved
ionization stages.  These corrections can be large and
uncertain when the unobserved ions constitute a significant
fraction of an element's overall abundance.  Second, while
robust ionization corrections can be derived from numerical
simulations of nebulae \cite{ferland98, kallman01}, this method
relies on the availability of accurate atomic data for
processes that affect the ionization equilibrium of each
element.  In photoionized nebulae, these atomic data include
photoionization cross sections and rate coefficients for
radiative and dielectronic recombination and charge exchange
reactions.  These data are unknown for the overwhelming majority of
\emph{n}-capture element ions. Uncertainties in the
photoionization and recombination data of \emph{n}-capture
element ions can result in elemental abundance uncertainties of
a factor of two or more \cite{sterling07}.

The present work is part of a larger study to determine the photoionization and recombination
properties of \emph{n}-capture element ions \cite{sterling_prep}, motivated by the astrophysical
detection of these species and the importance of measuring their elemental abundances accurately to test
theories of nucleosynthesis and
stellar structure.  The ultimate goal of this effort is to produce atomic data suitable for
incorporation into codes that numerically simulate the thermal and
ionization structure of nebulae, enabling significantly more accurate abundance
determinations of trans-iron elements in astrophysical nebulae than is presently achievable.  Determining these data over the range
of energies and temperatures encountered in
astrophysical environments necessitates a predominantly theoretical approach.  However,
experimental measurements are needed to constrain and establish the veracity of such
calculations, particularly in the case of complex systems such as low-charge states of
trans-iron elements.

Se was chosen as the first element of our investigation because it has been detected in
nearly twice as many planetary nebulae as any other trans-iron element \cite{sterling08}.
Experimental photoionization studies of other astrophysically observed \emph{n}-capture
elements have already been conducted by other groups for select Kr
\cite{lu06a, lu06b} and Xe ions \cite{bizau06, emmons05}.

This paper presents experimental determinations of the absolute Se$^+$
photoionization cross section near the ground-state ionization threshold, and is the first in
a series of papers on the photoionization of low-charge Se ions (up to 5 times ionized).  In
Section~2, the experimental procedure for our photoionization cross-section measurements is
described in detail.  The results and analysis of the data are presented in Section~3, and in
Section~4 we summarize our work.

\section{Experiment}

High-resolution measurements of the Se$^+$ photoionization cross section
have been carried out at the
Advanced Light Source (ALS) synchrotron radiation facility
at Lawrence Berkeley National
Laboratory in California. This experiment
used the merged beams technique~\cite{lyonBa+:jpb:86} with
the Ion Photon Beamline (IPB) apparatus located at undulator
beamline 10.0.1 of the ALS.  A detailed description of the IPB
apparatus is available in Covington \textit{et
al.}~\cite{covingtonNe+:pra:02}.  The IPB endstation has been used
for photoionization cross-section measurements of a variety of singly-
and multiply-charged
ions~\cite{mullerC2+:jpb:02,schippersB+:03,fred04,scully05,scully06,muller07}.

Se ions were produced by gently heating solid selenium inside a
resistive oven within an electron-cyclotron-resonance (ECR) ion
source.  These ions were accelerated to an energy of 6~keV, and
a 60$^{\circ}$ analyzing magnet selected Se$^+$ from the
accelerated ion beam.  The Se$^+$ ions were collimated with two
sets of vertical and horizontal slits and focused by three
electrostatic einzel lenses. The resulting collimated ion beam
had a typical diameter of a few millimeters, and a current
ranging from 20 to 200~nA. The ions were merged onto the axis
of the counter-propagating photon beam by a pair of
90$^{\circ}$ spherical bending plates. In the merged beam path,
an electrical potential of 1.4~kV was placed on the
``interaction region'' to energy-label photoions produced in a
well-defined volume, for the purpose of absolute
photoionization cross-section measurements. The interaction
region consists of an isolated stainless-steel mesh cylinder
with entrance and exit apertures defining an effective length
of 29.4 cm. Two-dimensional intensity distributions of the
photon and ion beams were measured by commercial rotating-wire
beam profile monitors installed on either side of the
interaction region, and by three translating-slit scanners
located within the cylinder.  Downstream from the interaction
region, the Se$^{2+}$ product ions were separated from the
parent Se$^+$ ion beam with a 45$^{\circ}$ dipole demerger
magnet.  This directed the Se$^+$ ions to a Faraday cup, while
the photoions were steered by a spherical 90$^{\circ}$
electrostatic deflector onto a negatively biased stainless
steel plate.  The secondary electrons produced by the Se$^{2+}$
collisions on this plate were recorded by a single-particle
channeltron detector. The detection efficiency
has been determined on several occasions by measuring a femtoampere 
ion current at the stainless steel plate and
comparing with the count rate generated from the
channeltron. These measurements have consistently shown 100\%
efficiency for this detection scheme.

Photons were produced by a 10-cm period undulator
located in the 1.9~GeV electron storage ring of the ALS.
A grazing-incidence spherical grating monochromator delivered a
highly collimated photon beam of spatial width less than 1 mm
and divergence less than 0.05$^{\circ}$.
The photon energies were selected and scanned
by rotating the grating and translating the exit slit
of the monochromator, while simultaneously adjusting the
undulator gap to maximize the beam intensity. The spectral resolution of
the photon beam was controlled with the entrance and exit slits
of the monochromator.  The photon flux was typically
3$\times$10$^{13}$ photons/sec, as measured by a silicon
x-ray photodiode (IRD, SXUV-100) that was referenced to two identical photodiodes absolutely
calibrated by the National Institute of Standards and Technology (NIST) and by the National Synchrotron Light Source (NSLS).  To calibrate the photon energy, the
well known doubly-excited states of He~\cite{domke96}
were measured in first, second and third order.
These measurements indicated that the uncertainty in the photon
energies of the reported cross sections can be conservatively estimated to be
less than 10~meV.  The photon beam was
mechanically chopped to separate photoions from the background
produced by collisions between the parent ion beam and residual
gas inside the interaction region.

\section{Results and discussion}

\begin{figure}[ht!]
\includegraphics[width=14cm,height=10cm]{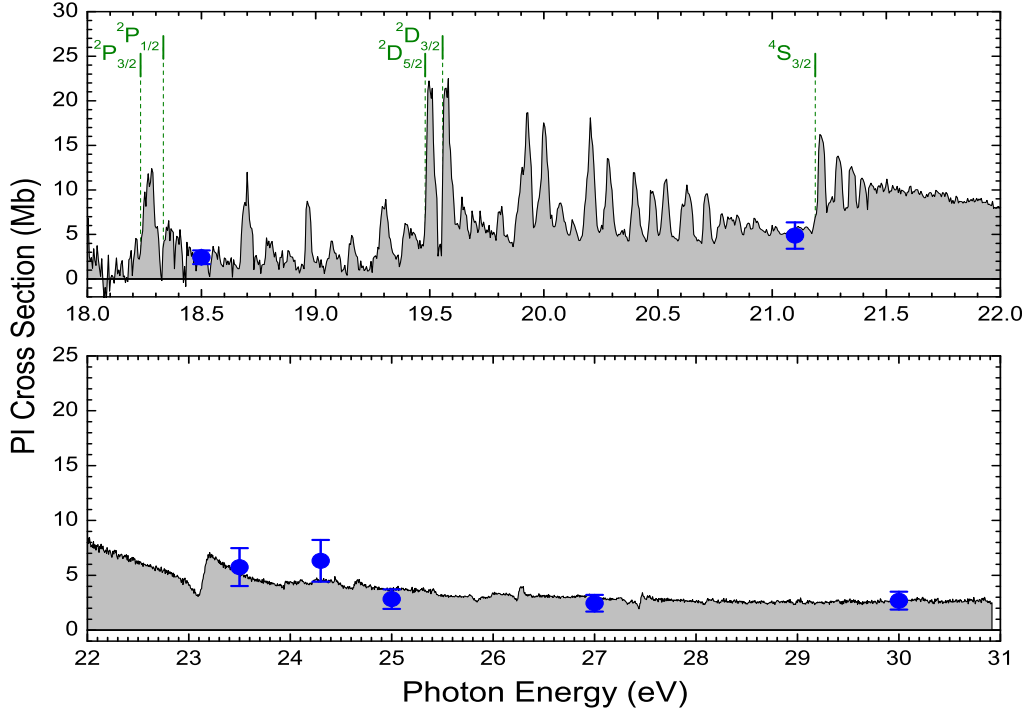}
\caption{Absolute cross-section measurements of photoionization of
Se$^{+}$ are shown from 18 eV to 31 eV, at 28~meV resolution. The ionization thresholds of the quartet ground state and the four doublet metastable states (from NIST \cite{NIST}) are indicated by vertical dashed lines. The shaded region represents energy scan measurements which are normalized to absolute cross-section measurements at specific energies (solid circles with error bars). \label{lores}}
  \end{figure}

\begin{figure}[t!]
\includegraphics[width=14cm,height=10cm]{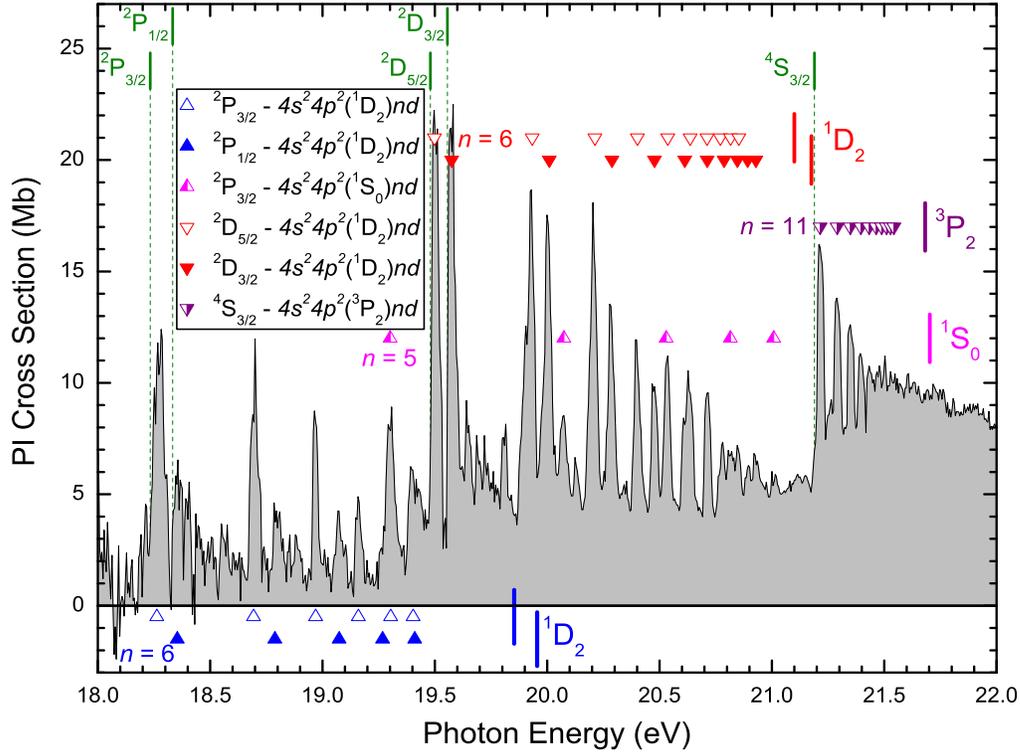}
\caption{Absolute photoionization cross-section measurements
in the region of the ground and metastable state thresholds at an energy resolution of 28 meV.
Resonances originating from the $^2P^o$ metastable state have been identified
as members of three different Rydberg series (open and solid triangles, and half-filled
pink triangles; see Table~\protect\ref{tableSe2P}). Two more Rydberg series
from the $^2D^o$ metastable state have been distinguished
(open and filled red triangles) and are listed in Table~\protect\ref{tableSe2D}.
All these series converge to either the $^1D$ or the $^1S$ excited
states of Se$^{2+}$ as indicated by thick vertical lines.
One series originating from the $^4S_{3/2}$ ground state converging to the
$^3P_2$ state of Se$^{2+}$ has also been identified and is indicated by
half-filled purple triangles.
The vertical dashed lines indicate the reported
metastable and ground-state ionization thresholds from NIST \cite{NIST}.\label{IDFig}}
  \end{figure}

The photoion yield for Se$^{+}$ was measured from 18 eV to 31
eV at a photon energy resolution of 28 meV (Figures~\ref{lores}
and \ref{IDFig}).  The actual resolution was determined by fitting Lorentzian
profiles to isolated features across the scanned energy region, and taking
the average value of the fitted resonance widths.
Absolute photoionization
cross sections were measured at discrete photon energies with
the same resolution.  The photoionization spectrum was multiplied by 
a polynomial function to normalize the spectroscopic data to all
absolute cross-section measurements, which are indicated in 
Figure~\ref{lores} by
solid circles with associated uncertainties.\footnote{All
uncertainties quoted in this paper are 90\% confidence level
estimates.}   Absolute photoionization measurements with the IPB
apparatus at the ALS typically are uncertain by $\sim$20\% at a
90\% confidence
level~\cite{covingtonNe+:pra:02,aguilarO+:apjs:03}.  For a
detailed discussion of uncertainty estimates for
photoionization measurements with the IPB apparatus, see
\cite{covingtonNe+:pra:02,aguilarO+:apjs:03}.  Note that these
uncertainties do not account for contamination from
higher-order radiation from the undulator at low photon
energies.  A previous experiment on Xe$^{3+}$
\cite{emmons_thesis} using the same beamline estimated the contamination of higher-order
radiation to be $\sim$2\% near 40~eV.  At lower photon energies, the contamination
of higher-order radiation is expected to be larger, but not by more than a factor of
2--3 compared to the contamination at 40~eV.  The total experimental uncertainties of 
the absolute measurements are estimated
to be 30\%, which accounts for the possible contamination of the photon
beam by higher-order radiation.  We note that the lowest-energy absolute cross
section values at 18.5 and 21.1~eV may not be as accurate as
the others due to the effects of higher-order radiation.



Given the similarities in electronic structure of Se$^+$ and O$^+$ --- both
have two $p$ electrons in their outermost shell that give rise to the same $^4S$
ground state and $^2P$ and $^2D$ metastable states --- their photoionization
spectra are expected to be similar.  Indeed, a comparison
between the measured Se$^+$ photoionization cross section and that of O$^+$ reported by Aguilar \emph{et al.}
\cite{aguilarO+:apjs:03} shows that to be the case.  For O$^+$, the region of the spectrum encompassing
the $^2P^o$ and $^4S^o$ thresholds consists of very strong resonances due to 2$p$ $\rightarrow$ $nd$ electron excitations of the metastable ions, as well as weak resonances due to 2$p$ $\rightarrow$ $ns$ transitions. At our resolution of 28 meV, only the 4$p$ $\rightarrow$ $nd$ transitions are observed in the spectrum of Se$^+$.


Identifications for the observed Se$^+$ structure (see Figure~\ref{IDFig})
are made based on the O$^+$ resonance identifications and the use of the quantum-defect form of the Rydberg formula,
 \begin{equation}
 E_n=E_{\infty} - \frac{R(Z-N)^2}{(n-{\delta}_n)^2}
 \label{Ryd_formula}
 \end{equation}
 where $Z$ is the charge of the nucleus, $N$ is the number
 of core electrons, $n$ is the principal quantum number, $E_{\infty}$
 is the series limit and ${\delta}_n$ is the dimensionless
 quantum defect parameter that indicates the departure of
 the energy level $E_n$ from the hydrogenic value.  Two series
 from the $^2P^o_{3/2}$ metastable state and one from
 $^2P^o_{1/2}$ have been identified.  The 4$s^2$4$p^2$($^1D$)$nd$
 series converging to the $^1D$ series limit is shown in
 Figure~\ref{IDFig} by open blue triangles for resonances
 originating from the $^2P^o_{3/2}$ state, and by filled blue
 triangles for resonances from $^2P^o_{1/2}$.
 The 4$s^2$4$p^2$($^1S$)$nd$ series converging to the
 $^1S$ series limit originating from the $^2P^o_{3/2}$ state
 is depicted by half-filled pink triangles in Figure~\ref{IDFig}.
 In addition, two other series from the $^2D^o_j$ ($j$ = 5/2, 3/2)
 metastable states are indicated by inverted open and filled red
 triangles above the spectrum. These resonances correspond to
 the 4$s^2$4$p^2$($^1D$)$nd$ series converging to the $^1D$ limit.
 In the measured energy range, only one Rydberg series is
 observed from the $^4S_{3/2}$ ground state, whose first
 autoionizing member is 4$s^2$4$p^2$($^3P_2$)$11d$. The
 series is indicated with inverted half-open purple triangles
 just above the ground state threshold.

 Tables~\ref{tableSe2P},~\ref{tableSe2D} and~\ref{tableSe4S}
 list the principal quantum numbers, resonance
 energies and quantum defects of the identified members of each series.
The uncertainty in the quantum defect values of
the first few resonances in each series is conservatively
estimated to be 10\%. These uncertainties are a
function of the energy uncertainty as well as the relative
precision to which individual resonances can be identified and
their associated centroids resolved. High $n$-value resonances
are typically much more difficult to clearly identify as they
have significantly lower intensities and are often obscured by
strong low-$n$ resonances in adjacent series. High-$n$ resonances
can also become unresolvable as series converge toward their
respective limits. Due to these complications, uncertainties have not been
estimated for high-$n$ resonances.

It is important to note that ECR ion sources are known to
produce ions in the ground and metastable states in fractions
that may differ from statistically-weighted values. Therefore,
the reported cross-section measurements correspond to an
unknown admixture of metastable and ground state fractions. In
the case of O$^+$~\cite{aguilarO+:apjs:03} the metastable
fractions were determined using the beam attenuation method
($^4S$ 43\%, $^2D$ 42\% and $^2P$ 15\%), clearly differing from
the statistically-weighted values ($^4S$ 20\%, $^2D$ 50\% and
$^2P$ 30\%) but not as much as those reported using
translational energy spectroscopy ($^4S$ 60\%, $^2D$ 19\% and
$^2P$ 21\%)~\cite{Enos_fractions:jpb:92}.  Similarly, we do not
expect statistically-weighted metastable fractions in our Se$^+$
measurements.

\begin{table}
\centering
\begin{minipage}{15.0cm}
\centering
\addtocounter{footnote}{1}
\caption{Principal quantum numbers ($n$), resonance
      energies and quantum defects ($\delta$) determined from
      the present measurements of photoionization of Se$^+$. Two distinct Rydberg series are observed
      from the $^2P^{o}_{3/2}$ and one from the $^2P^{o}_{1/2}$ metastable states of Se$^+$, all due to $4p \rightarrow nd$ transitions. The uncertainties in
      the experimental energies are estimated to be $\pm$0.010 eV or less.}\label{tableSe2P}
\begin{tabular}{cccc@{\ \ \ \ \ \ \ \ \ }cccc@{\ \ \ \ \ \ \ \ \ }ccccc}
\hline\hline\\
\multicolumn{8}{c}{\bfseries Initial Se$^+$ state: $\:4s^24p^3\ (^2P^{o}_{3/2})$}&&&\\
 \hline
\multicolumn{3}{c}{Rydberg Series}&&&\multicolumn{3}{c}{Rydberg Series} &&& \\[.02in]
\multicolumn{3}{c}{\small{$\:4s^24p^2(^1D_2)nd $}}&&&\multicolumn{3}{c}{\small{$\:4s^24p^2(^1S_0)nd$}}&&&\\
$n$&Energy (eV)&$\delta$&&&$n$&Energy (eV)&$\delta$&&&&&\\[.02in]
\hline\\
6 &  18.268 & 0.14 &&& 5 & 19.301 & 0.24 &&&&&\\
7 &  18.697 & 0.14 &&& 6 & 20.074 & 0.22 &&&&&\\
8 &  18.972 & 0.14 &&& 7 & 20.530 & 0.19 &&&&&\\
9 &  19.160 & 0.14 &&& 8 & 20.815 & 0.17 &&&&&\\
10 & 19.296 & 0.12 &&& 9 & 21.008 & 0.15 &&&&&\\
11 & 19.395 & 0.1    &&& - &    -   &   -  &&&&&\\
$\cdot$ & $\cdot$ &- &&& $\cdot$ & $\cdot$ &&&&&&\\
$\infty$ & 19.853\footnote{NIST tabulations~\cite{NIST}} &- &&& $\infty$ & 21.703\footnotemark[1] & - &&&\\
\hline\hline\\
\multicolumn{3}{c}{\bfseries Initial Se$^+$ state: $\:4s^24p^3\ (^2P^{o}_{1/2})$}\\
 \hline
\multicolumn{3}{c}{Rydberg Series} \\[.02in]
\multicolumn{3}{c}{\small{$\:4s^24p^2(^1D_2)nd$}}\\
$n$&Energy (eV)&$\delta$\\[.02in]
\hline\\
6  & 18.354 & 0.17 \\
7  & 18.788 & 0.17 \\
8  & 19.070 & 0.16 \\
9  & 19.262 & 0.14 \\
10 & 19.298 & 0.12    \\
$\cdot$ & $\cdot$ & - \\
$\infty$ & 19.955\footnotemark[1] & - \\
\hline\hline\\
\end{tabular}
\end{minipage}
\end{table}

\begin{table}
\centering
\begin{minipage}{15.0cm}
\centering
\addtocounter{footnote}{1}
\caption{Principal quantum numbers ($n$), resonance
      energies and quantum defects ($\delta$) determined from
      the present measurements of photoionization of Se$^+$. Two distinct Rydberg series are observed
      from the $^2D^o$ metastable state of Se$^+$, due to $4p \rightarrow nd$
      transitions. The uncertainties in
      the experimental energies are estimated to be $\pm$0.010 eV or less.}\label{tableSe2D}
\begin{tabular}{cccc@{\ \ \ \ \ \ \ \ \ \ \ \ }cccc}
\hline\hline\\
\multicolumn{3}{c}{\bfseries Initial Se$^+$ state: $\:4s^24p^3\ (^2D^{o}_{5/2})$}&&&
\multicolumn{3}{c}{\bfseries Initial Se$^+$ state: $\:4s^24p^3\ (^2D^{o}_{3/2})$}\\
 \hline
\multicolumn{3}{c}{Rydberg Series}&&&\multicolumn{3}{c}{Rydberg Series}\\[.02in]
\multicolumn{3}{c}{\small{$\:4s^24p^2(^1D_2)nd $}}&&&\multicolumn{3}{c}{\small{$\:4s^24p^2(^1D_2)nd$}}\\
$n$&Energy (eV)&$\delta$&&&$n$&Energy (eV)&$\delta$\\[.02in]
\hline\\
6 &  19.499 & 0.17 &&&  6  & 19.575 & 0.17\\
7 &  19.933 & 0.17 &&&  7  & 20.009 & 0.17 \\
8 &  20.212 & 0.17 &&&  8  & 20.288 & 0.17 \\
9 &  20.402 & 0.17 &&&  9  & 20.478 & 0.17 \\
10 & 20.537 & 0.17 &&&  10 & 20.613 & 0.17 \\
11 & 20.637 & 0.16 &&&  11 & 20.713 & 0.16\\
12 & 20.712 & 0.16 &&&  12 & 20.788 & 0.16 \\
13 & 20.770 & 0.15 &&&  13 & 20.846 & 0.15 \\
14 & 20.816 & 0.15 &&&  14 & 20.892 & 0.15 \\
15 & 20.853 & 0.15 &&&  15 & 20.929 & 0.15 \\
$\cdot$ & $\cdot$ & - &&& $\cdot$ & $\cdot$ & - \\
$\infty$ & 21.100\footnote{NIST tabulations~\cite{NIST}} & - &&& $\infty$ & 21.167\footnotemark[1]& - \\
\hline\hline\\
\end{tabular}
\end{minipage}
\end{table}

\begin{table}
\centering
\begin{minipage}{15.0cm}
\centering
\addtocounter{footnote}{1}
\caption{Principal quantum numbers ($n$), resonance
      energies and quantum defects ($\delta$) determined from
      the present measurements of photoionization of Se$^+$. One distinct Rydberg series is observed
      from the $^4S^o$ metastable state of Se$^+$, due to $4p \rightarrow nd$ transitions. The uncertainties in
      the experimental energies are estimated to be $\pm$0.010 eV or less.}\label{tableSe4S}
\begin{tabular}{ccc}
\hline\hline\\
\multicolumn{3}{c}{\bfseries Initial Se$^+$ state: $\:4s^24p^3\ (^4S^o_{3/2})$}\\
 \hline
\multicolumn{3}{c}{Rydberg Series}\\[.02in]
\multicolumn{3}{c}{\small{$\:4s^24p^2(^3P_2)nd $}}\\
$n$&Energy (eV)&$\delta$\\[.02in]
\hline\\
11 &  21.215 & 0.21 \\
12 &  21.291 & 0.21 \\
13 &  21.349 & 0.21 \\
14 &  21.396 & 0.21 \\
15 &  21.433 & 0.21 \\
16 &  21.464 & 0.21 \\
17 &  21.489 & 0.21 \\
18 &  21.510 & 0.21 \\
19 &  21.528 & 0.21 \\
20 &  21.543 & 0.21 \\
$\cdot$ & $\cdot$ &- \\
$\infty$ & 21.682\footnote{NIST tabulations~\cite{NIST}} &- \\
\hline\hline\\
\end{tabular}
\end{minipage}
\end{table}


\section{Summary}

The absolute photoionization cross section of Se$^+$ has been measured
in the energy region of the ground state ionization threshold.  The cross section exhibits a wealth of
resonances that form a clear pattern of Rydberg series.  The strongest resonances are
identified as 4$p$$\rightarrow$$nd$ transitions belonging to the  4$s^2$4$p^2$($^1$D$_2$)$nd$
and 4$s^2$4$p^2$($^1$S$_0$)$nd$ series originating from the $^2P^o$ and $^2D^o$ metastable states.
The sole series from the $^4S_{3/2}$ ground state is identified as 4$s^2$4$p^2$($^3P_2$)$nd$.
The resonance positions and quantum defects are
determined for the initial members of each of these series.

The photoionization cross sections from the present study (and for other low-charge
Se ions, to be presented in forthcoming papers) will be used to calibrate a
broader theoretical effort to determine the photoionization and recombination
properties of astrophysically observed \emph{n}-capture elements \cite{sterling_prep}.
The resulting atomic data determinations will enable the abundances of trans-iron species
in astrophysical nebulae to be derived to a much higher degree of accuracy than
is currently possible, which bears implications for the nucleosynthetic sites and chemical
evolution of Se and other trans-iron elements.  Our absolute cross sections for Se$^+$ can be
accessed via secure FTP at
the IP address 131.243.76.25 (note that a username and password are required; these can be
obtained by contacting A Aguilar at aaguilar@lbl.gov), or by contacting the authors
N C Sterling (sterling@pa.msu.edu) and A Aguilar.

\ack

We acknowledge support by the Director, Office of Science, Office of Basic Energy Sciences, of the U.S. Department of Energy under contracts DE-AC02-
05CH11231, DE-AC03-76SF-00098, and grant DE-FG02-03ER15424. N C Sterling acknowledges support from an NSF Astronomy and Astrophysics Postdoctoral Fellowship under award AST-0901432 and from NASA grant 06-APRA206-0049.  D.\ Esteves acknowledges the support from the Doctoral Fellowship Program at the Advanced Light Source. We thank Dr.\ Jeff Keister from Brookhaven National Laboratory and Dr.\ Robert Vest from NIST for performing absolute calibrations of the photodiodes.

\bibliographystyle{unsrt_notitle}

\bibliography{sterling_se+}

\begin{thebibliography}{10}

\bibitem{sterling07}
{Sterling N C, Dinerstein H L and Kallman T R}.
\newblock {\em {Astrophys.\ J.\ Suppl.\ Ser.}}, \textbf{169}:37, 2007.

\bibitem{sterling08}
{Sterling N C and Dinerstein H L}.
\newblock {\em {Astrophys.\ J.\ Suppl.\ Ser.}}, \textbf{174}:157, 2008.

\bibitem{smith90}
{Smith V V and Lambert D L}.
\newblock {\em {Astrophys.\ J.\ Suppl.\ Ser.}}, \textbf{72}:387, 1990.

\bibitem{wally97}
{Wallerstein G \etal}.
\newblock {\em {\RMP}}, \textbf{69}:995, 1997.

\bibitem{busso99}
{Busso M, Gallino R and Wasserburg G J}.
\newblock {\em {Ann.\ Rev.\ Astron.\ Astrophys.}}, \textbf{37}:239, 1999.

\bibitem{trav04}
{Travaglio C, Gallino R, Arnone E, Cowan J J, Jordan F and Sneden C}.
\newblock {\em {Astrophys.\ J.}}, \textbf{601}:864, 2004.

\bibitem{herwig05}
{Herwig F}.
\newblock {\em {Ann.\ Rev.\ Astron.\ Astrophys.}}, \textbf{43}:435, 2005.

\bibitem{sneden08}
{C.\ Sneden, J.\ J.\ Cowan, and R.\ Gallino}.
\newblock {\em {Ann.\ Rev.\ Astron.\ Astrophys.}}, \textbf{46}:241, 2008.

\bibitem{karakas09}
{Karakas A I, van~Raai M A, Lugaro M, Sterling N C and Dinerstein H L}.
\newblock {\em {Astrophys.\ J.}}, \textbf{690}:1130, 2009.

\bibitem{pequignot94}
{P\'{e}quignot D and Baluteau, J P}.
\newblock {\em {Astron.\ Astrophys}}, \textbf{283}:593, 1994.

\bibitem{sharpee07}
{Sharpee B, Zhang Y, Williams R, Pellegrini E, Cavagnolo K, Baldwin J A,
  Phillips M and Liu, X-W}.
\newblock {\em {Astrophys.\ J.}}, \textbf{659}:1265, 2007.

\bibitem{sterling09}
{Sterling N C \etal}.
\newblock {\em {Pub.\ Astron.\ Soc.\ Australia}}, \textbf{26}:339, 2009.

\bibitem{stran06}
{Straniero O, Gallino R and Cristallo S}.
\newblock {\em {Nucl.\ Phys.\ A}}, \textbf{777}:311, 2006.

\bibitem{karakas10}
{Karakas A I and Lugaro M}.
\newblock {\em {Pub.\ Astron.\ Soc.\ Australia}}, \textbf{27}:227, 2010.

\bibitem{ferland98}
{Ferland G J, Korista K T, Verner D A, Ferguson J W, Kingdon J B and Verner E
  M}.
\newblock {\em {Pub.\ Astrophys.\ Soc.\ Pac.}}, \textbf{110}:761, 1998.

\bibitem{kallman01}
{Kallman T and Bautista M}.
\newblock {\em {Astrophys.\ J.\ Suppl.\ Ser.}}, \textbf{133}:221, 2001.

\bibitem{sterling_prep}
{Sterling N C and Witthoeft M C}.
\newblock {\em {Astron.\ Astrophys.}}, {in preparation}.

\bibitem{lu06a}
{Lu M \etal}.
\newblock {\em {Phys.\ Rev.\ A}}, \textbf{74}:012703, 2006.

\bibitem{lu06b}
{Lu M, Alna'Washi G, Habibi M, Gharaibeh M F, Phaneuf R A, Kilcoyne A L D,
  Levenson E, Schlacter A S, Cisneros C and Hinojosa G}.
\newblock {\em {Phys.\ Rev.\ A}}, \textbf{74}:062701, 2006.

\bibitem{bizau06}
{Bizau J M, Blancard C, Cubaynes D, Folkmann F, Champeaux J P, Lemaire J L and
  Wuilleumier F J}.
\newblock {\em {Phys.\ Rev.\ A}}, \textbf{73}:022718, 2006.

\bibitem{emmons05}
{Emmons E D, Aguilar A, Gharaibeh M F, Scully S W J, Phaneuf R A, Kilcoyne A L
  D, Schlachter A S, \'{A}lvarez I, Cisneros C and Hinojosa G}.
\newblock {\em {Phys.\ Rev.\ A}}, \textbf{71}:042704, 2005.

\bibitem{lyonBa+:jpb:86}
{Lyon I C, Peart B, West J B and Dolder K}.
\newblock {\em {\jpb}}, \textbf{19}:4137, 1986.

\bibitem{covingtonNe+:pra:02}
{Covington A M \etal}.
\newblock {\em {Phys.\ Rev.\ A.}}, \textbf{66}:062710, 2002.

\bibitem{mullerC2+:jpb:02}
{M\"{u}ller A, Phaneuf R A, Aguilar A, Gharaibeh M F, Schlachter A S, Alvarez
  I, Cisneros C, Hinojosa G and McLaughlin B M}.
\newblock {\em {\jpb}}, \textbf{35}(7):L137, 2002.

\bibitem{schippersB+:03}
{Schippers S, M\"{u}ller A, McLaughlin B M, Aguilar A, Cisneros C, Emmons E,
  Gharaibeh M F and Phaneuf R A}.
\newblock {\em {\jpb}}, \textbf{36}:3371, 2003.

\bibitem{fred04}
{Schlachter A S \etal}.
\newblock {\em {\jpb}}, \textbf{37}:L103, 2004.

\bibitem{scully05}
{Scully S W J \etal}.
\newblock {\em {\jpb}}, \textbf{38}:1967, 2005.

\bibitem{scully06}
{Scully S W J \etal}.
\newblock {\em {\jpb}}, \textbf{39}:3957, 2006.

\bibitem{muller07}
{M\"{u}ller A, Schippers S, Phaneuf R A, Kilcoyne A L D, Br\"{a}uning H,
  Schlachter A S, Lu M and McLaughlin B M}.
\newblock {\em {J.\ Phys.\ Conf.\ Ser.}}, \textbf{58}:383, 2007.

\bibitem{domke96}
{Domke M, Schulz K, Remmers G and Kaindl G}.
\newblock {\em {Phys.\ Rev.\ A}}, \textbf{53}:1424, 1996.

\bibitem{aguilarO+:apjs:03}
{Aguilar A \etal}.
\newblock {\em {Astrophys.\ J.\ Suppl.\ Ser.}}, \textbf{146}:467, 2003.

\bibitem{emmons_thesis}
{Emmons E D}.
\newblock Master's thesis, {University of Nevada, Reno}, 2004.

\bibitem{NIST}
{Ralchenko Y, Kramide A E, Reader J and NIST ASD Team}.
\newblock {National Institute of Standards and Technology, Gaithersburg, MD.},
  2008.
\newblock {http://physics.nist.gov/cgi-bin/AtData/main\_asd3}.

\bibitem{Enos_fractions:jpb:92}
{Enos C S, Lee A R and Brenton A G}.
\newblock {\em {\jpb}}, \textbf{25}:4021, 1992.

\end{thebibliography}

\end{document}